# Acoustic Emission Cascade Predicting Progression to Failure of Rock and Bone

Andrew P. Bunger[a,b,*], Yunxing Lu[a], Ayyaz Mustafa[a], Michael M. McDowell[c]

a)  Department of Civil and Environmental Engineering, Swanson School of Engineering, University of Pittsburgh, Pittsburgh, PA, USA
b)  Department of Chemical and Petroleum Engineering, Swanson School of Engineering, University of Pittsburgh, Pittsburgh, PA, USA
c)  Department of Neurosurgery, Penn State Health Children's Hospital and Penn State College of Medicine, Hershey, Pennsylvania, USA

* Corresponding author: bunger@pitt.edu. 710 Benedum Hall, 3700 O'Hara Street, Pittsburgh, PA, 15261

# Abstract

Quasi brittle materials such as rock and bone are understood to fail via microcrack coalescence. The accompanying Acoustic Emission (AE) event rate is known to increase as failure progresses. Here we examine the progression of the AE event rate for both rock and bone under conditions where failure progresses under fixed loading. The experiments for rock entail subjecting granite beams to a fixed loading under three point bending and monitoring the time-dependent failure. For bone, human cadaver skulls are loaded under pinning loads similar to those used for immobilization of the head for neurosurgical procedures. AE rates are shown to be consistent with a rate dependent material failure law including a quasi-linear cascade of the inverse AE energy rate in the lead up to failure that is shown here to arise because of a coupling wherein a response rate is a power law of a driver and the driver is, in turn, a power law of the accumulated response. For both materials and experimental configurations this cascade provides warning of impending failure. For granite beams there is accurate prediction of failure time over the final 30% of the specimens' lifetimes, which ranged from 35 seconds to two days. For skull fracture, the commencement of the failure cascade provides 30-70 seconds for warning of failure, which would be sufficient for basic remedial action to avoid patient injury in a surgical application.

# Introduction

Rate-dependent failure of materials under constant loading has been widely observed to follow an empirical relationship between acceleration and velocity of some observable quantity related to the progression toward failure, $\Omega$, given by (Lei and Sornette 2024 after Voight 1988)

$$\ddot{\Omega} = \eta \dot{\Omega}^{\alpha} \qquad (1)$$

Here the overdot indicates differentiation with respect to time and $\alpha$ and $\eta$ are obtained to fit experimental data. Typically $\Omega$ is taken as material strain, displacement, energy release, or similar. Originally proposed as a means for predicting volcanic eruptions based on field measurement data (Voight 1988), it has remained an important basis for various types of natural disaster forecasting including landslides, rockbursts, and volcanic eruptions (Lei and Sornette 2024). However, Voight (1989) demonstrated it to apply to a much wider range of materials, finding it to have "striking [but] incompletely understood generality" for the latter stages of rate-dependent material behavior under constant load. Its main feature is divergence of the measurable quantity to infinity if $\alpha > 1$, thereby indicating catastrophic failure at a critical time $t_c$ (Voight 1989). This trait is readily seen upon integration of Eq. (1) to obtain

$$\dot{\Omega} = \kappa(t_c - t)^{-\xi}, \quad \kappa = (\xi/\eta)^\xi, \quad \xi = 1/(\alpha - 1) \tag{2}$$

Voight (1989) finds $\alpha \approx 2$ for metals and soils and notes it is also the value that gives dimensional consistency.

Some of the most successful predictive applications have used a measurable displacement for the quantity $\Omega$. These include predictions of eruptions of Mt Saint Helens (e.g. Swanson et al. 1983) that precede publications of Voight's theory but are referenced and reanalyzed in Voight's later work (Voight 1988). Indeed, an accelerating displacement under fixed load is direct evidence of material softening in the leadup to failure. There have also been multiple studies using a measure of seismic energy as the measurement quantity $\Omega$. Voight (1988) includes one seismicity-related quantity in the analysis of prediction of Mt Saint Helens eruptions (the square root of the cumulative seismic energy) and the results are promising. Additionally, seismicity building is shown concurrently with deformation measurements that are complementary for eruption prediction by Voight et al. (2000). However, in other settings, precursors to failure that can be used to generate robust predictions of eruptions are less consistently observed (e.g. Chastin and Main 2003). With that said, Chastin and Main (2003) point out that confounding factors are likely related to geologic complexity and not to fundamentals of rock failure.

On the other hand, it is reasonable to hypothesize that time-dependent rock failure under fixed loading conditions in the laboratory should be accompanied by a cascade of acoustic emission (AE) owing to the underlying microcrack coalescence mechanism of failure. Indeed, Winner et al (2018) show a period of slow acceleration followed by rapid acceleration of AE events during the final roughly half of the time beams of granite are able to sustain constant load level that is below the load required to induce immediate failure but sufficient to induce time-delayed failure.

The first focus of this paper is therefore to present a reanalysis of the AE data from Winner et al. (2018) in order to determine: 1) if the final cascade of AE preceding failure conforms to Voight's rate dependent material failure law, Eq (1) during the slow acceleration period, rapid acceleration period, neither, or both, and 2) if analysis of the final cascade can be used to generate long-range forecasting of the time of failure for granite beams.

For context, there is an existing body of work in the rock mechanics literature that uses the rate dependent failure law of Voight (1989) with AE event rates and/or AE energy rates to predict time to failure (Fan et al. 2024, Zhang et al. 2023, Wang et al. 2022, Zhang et al. 2020, Xue et al. 2018). However, the present work differs in one very important sense, namely, that the AE rate and progression to failure are measured under a constant load. This loading conforms to the assumptions originally presented for the theory (Voight 1989). The works to date in rock mechanics have considered non-constant load. In most cases the load was non-constant and associated with a constant applied strain rate (Fan et al. 2024). The utility of the approach for predicting the failure time is clear in these past contributions, but the time dependence of the AE rate or AE energy convolves both the rate dependence of the material failure itself and the transient nature of the loading. Here we consider a constant load so that the evolution of the AE energy rate can be considered to arise only from the rate dependence of the material failure. Note that AE detection under constant loading is previously considered by several prior authors (Heap et al. 2009, Rudajev et al. 2000, Mogi 1962), and broad similarity can be seen in the event rate to those obtained by Winner et al. (2018). The availability of the full data set for Winner et al. (2018) makes it possible to reanalyze these data.

Such an exploration of AE precursors to the transition from stable creep to catastrophic failure under constant loading is motivated by a wide range of applications such a result could have in structural health monitoring. As one novel biomedical example of structural health monitoring, we then explore the potential to use AE failure cascade to prevent patient injuries during neurosurgical procedures for which secure fixation of the head is essential to safety. Fixation almost always uses a clamp that pushes three pins into the skull and that, in turn, can be affixed to the surgical table (Figure 1). Safe pinning balances using too much force and inducing skull fracture secondary to pinning with too little force and risking scalp lacerations and other injuries resulting from head slippage (e.g. Shah et al. 2023). While fracture and slippage can be problematic for patients of all ages, the most significant challenge is for pediatric patients, where the immature skull can leave a very narrow window of pinning forces that are sufficient to avoid slippage while not inducing skull fractures (e.g. Berry et al. 2008). Recent survey data from pediatric neurosurgeons indicates that skull fracture and slippage in pediatric neurosurgeries is significantly underreported and that methods and/or devices to approve patient safety are needed (McDowel et al. In Press).

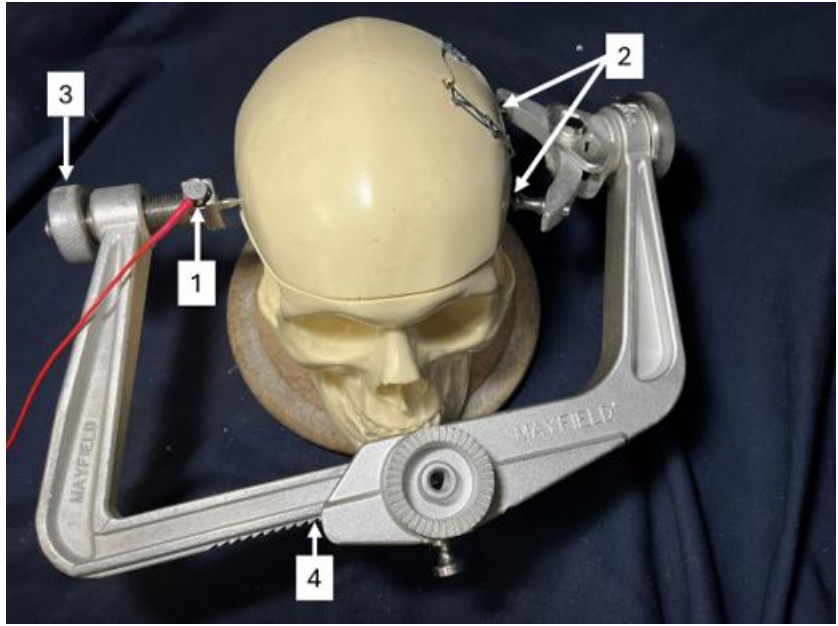

*Figure 1: Skull clamp attached to model of a human skull showing: 1) Single pin side, here with a modified pin that enables coupling to an Acoustic Emission (AE) transducer, 2) Two-pin side, 3) Screw drive that is turned to increasing pinning load, 4) Ratchet mechanism used for coarse adjustment upon initial attachment. After Bunger et al. (2024a, In Review).*

Specifically, we will examine the hypothesis that bone will fail in a time-dependent manner and with an AE failure cascade similar to rock. This is a reasonable hypothesis owing to the fact that bone and rock are both known to be quasi-brittle materials with a mechanism of microcrack coalescence (e.g. Danova et al. 2003, Zioupos and Curry 1994 for bone and Sarfarazi and Haeri 2016 for concrete and rock). In this broader context, AE associated with fracture of rock and concrete have been extensively studied (e.g. reviews of Ono 2018, Behnia et al. 2014, and Manthei and Plenkers 2018). However, it remains to demonstrate whether the acceleration of the AE rate in either material can be used to predict time to failure. Thus motivated, we present reanalysis of AE data for skull fracture in the cadaver study of Bunger et al. (2024a, 2024b, In Review) in order to determine: 1) if the final cascade of AE conforms to the rate dependent failure mode of Eq. (1), and 2) if analysis of the final cascade can provide sufficient warning of impending fracture that a surgical team could take remedial action to avoid patient injury.

The paper begins with experimental methods and a summary of previously-presented results for both rock and bone. The granite beam experiments are detailed previously in Winner et al. (2018) and the cadaver experiments on skull fracture under pinning loads are detailed previously by Bunger et al. (2024a, 2024b, In review). However, for completeness, these will be briefly recounted. The paper will then review the rate dependent failure theory, providing a discussion of its applicability and details of the manner in which the quantities needed to apply the theory are extracted from AE data. The results section presents evidence for conformance to critical state theory (Eq. 1) in the leadup to failure followed by application of the so-called inverse rate method to prediction of time to failure for all experimental scenarios. The paper concludes with a discussion of the basic components of rate-dependent material behavior, as captured by a rather wide class of rate dependent material models, that leads to the feedback loop that is the driver of criticality and catastrophic failure.

## Experimental Methods

### Granite Beams

Granite beams were prepared from Cold Spring Charcoal Granite to a size of 25.4 x 25.4 x 127 mm (1x1x5 inches). They were supported from below by two rods with a span of 101.6 mm (4 inches) and loaded from above by a center rod, as shown in Figure 2. The acoustic emission data was collected using a MISTRAS Express-8 system and ten Physical Acoustics Mini30S transducers, which have an active sensing diameter of 0.26 inches, an operating frequency range of 270 to 970 kilohertz, and a resonant frequency of 325 kilohertz.

The load was applied at a rate of 60 N/s and can be related to the maximum tensile stress generated at the bottom of the specimen at its center, $\sigma_t$, by

$$\sigma_t = \frac{3PL}{2bh^2}$$

Here, *P* is the applied load, *L* is the support span, and *b* and *h* are the width and height of the specimen. The loading rate of 60 N/s for these specimens corresponds to a nominal tensile stress rate of about 3 MPa/s, allowing the largest nominal tensile stresses—ranging from about 16.3 to 19.1 MPa—to be reached within a few seconds. The applied load was then held constant until failure, with the time to breakage varying from about 35 seconds to over 180,000 seconds (~50 hours). The details of the experiments and the material properties for these granite beams can be found in Winner et al. (2018).

Six experiments were performed, and they showed very similar behavior to one another in spite of having different applied stress levels and correspondingly vastly different times to failure. A notable feature, observed in this example and across all six beam tests, is the acceleration of the cumulative events in the approach to failure, as shown in Figure 3a. During this period, the event locations concentrate around the eventually observed fracture plane, as shown in Figure 3b.

In all six cases, Winner et al. (2018) indicate that the event rate decreases over the first roughly 40% of the lifetime of the specimen, as indicated in the representative example of

Figure 4. Then, somewhere between 40 and 50 percent of the time to failure, the event rate begins to gradually increase until there is a final burst at the very end. Winner et al. (2018) identify these as four phases of the failure process, with phases 3 and 4 comprising these final two phases preceding failure (Figure 4). Note these AE event rates bear some similarity to the strain evolution during creep tests under fixed loading (e.g. Hao et al. 2014). Here we will re-evaluate the data during these periods to see if they conform to critical state theory for rate-dependent material failure. We will also explore whether this model can be used to predict the time to failure.

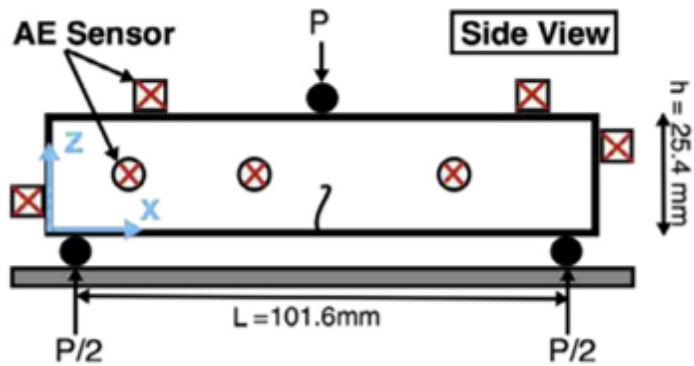

*Figure 2: Beam experiment setup, after Winner et al. (2018), used with permission.*

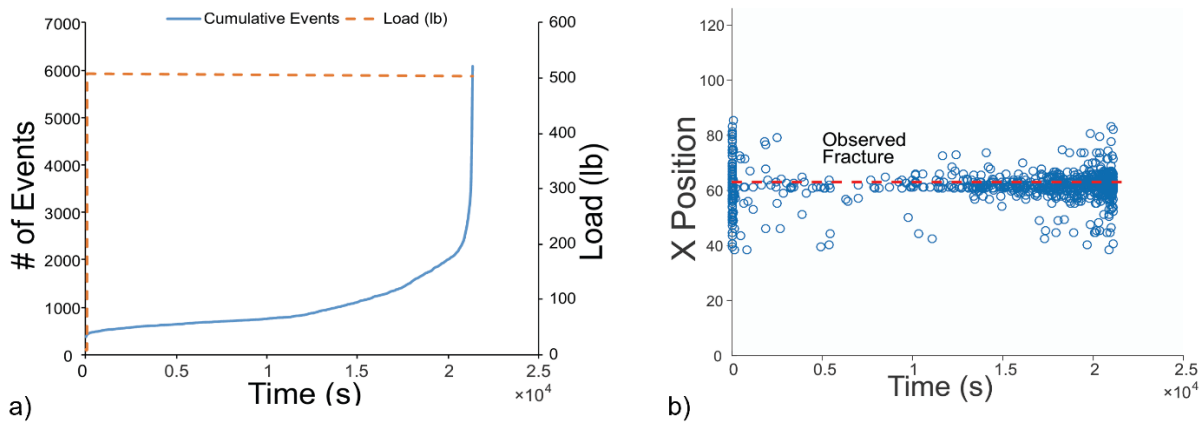

*Figure 3: Beam experiment example from a test lasting ~6 hrs, after Winner et al. (2018), used with permission, showing: a) Cumulative AE events as a function of time, and b) Location relative to eventual fracture surface as a function of time.*

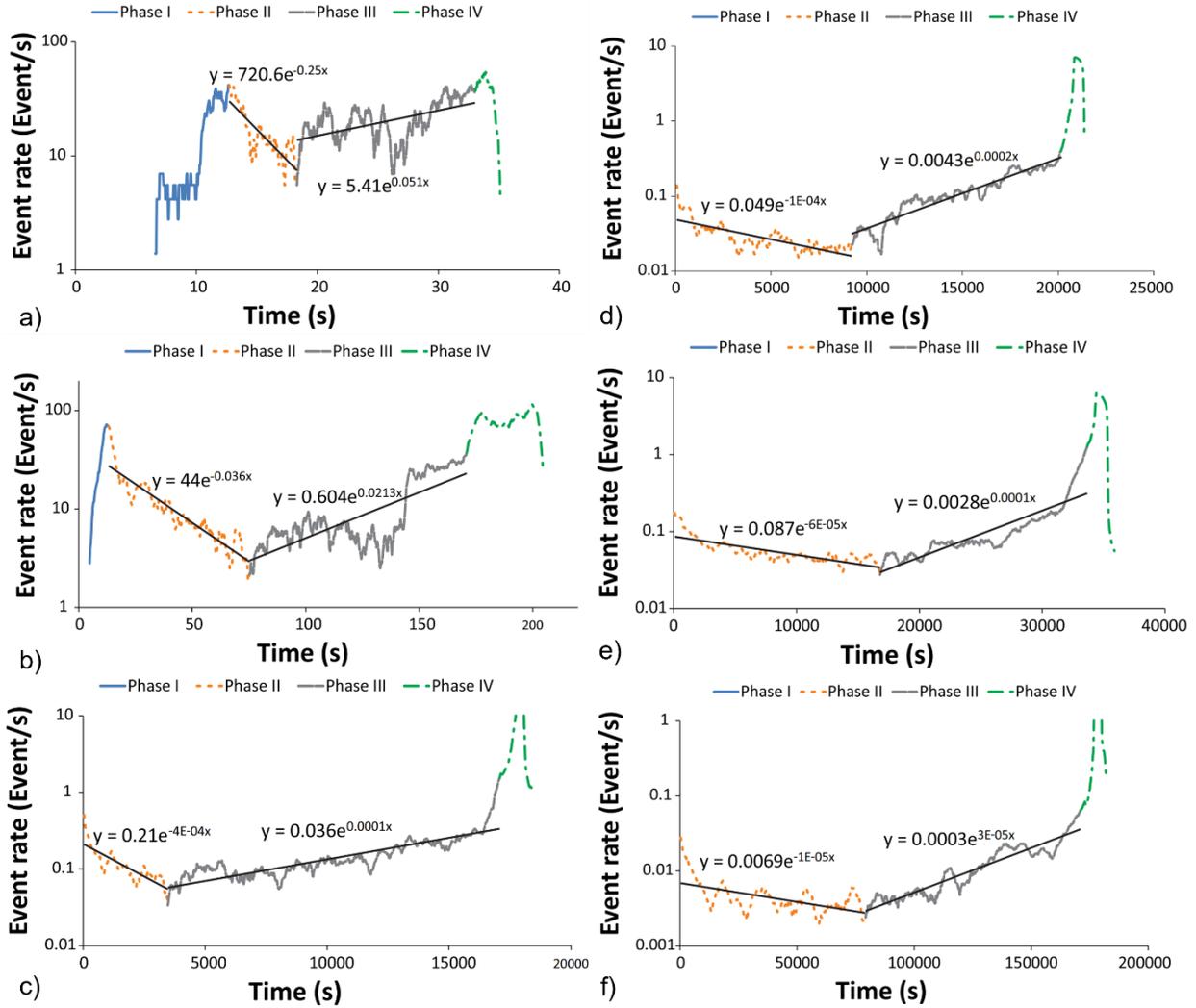

*Figure 4: All event rates, after Winner et al. (2018), with permission, showing cases with duration of: a) 35 seconds (~0.5 min), b) 205 sec (~3.4 min), c) 18,000 sec (~5 hr), d) 21,300 sec (~6 hr), e) 35,389 sec (~ 10 hr), f) 180,224 sec (~50 hr). For brevity these will henceforth be referred by to the abbreviate times the failure 0.5 min, 3.4 min, 5 hr, 6 hr, 10 hr, and 50 hr, respectively.*

## Skull Fracture under Pinning Loads

A piezoelectric acoustic emission (AE) transducer was mounted and acoustically coupled with a thin glucose layer to a modified pin attached to the screw drive on the single-pin side of a standard neurosurgical skull clamp. As with the granite beam tests, the AE transducer used was the Physical Acoustics Mini30S. An Integra LifeSciences Mayfield® clamp was used for these experiments, but other experiments (not presented here) were performed with the Black Forest Medical Group DORO® clamp. Both devices share a similar configuration: a single pin on the screw-drive side and a pair of pins with a rocker mechanism on the opposing side, as shown in Figure 1.

The clamp was secured to the frontotemporal region of a latex-injected, formalin-fixed, adult cadaver head. Fracture experiments were conducted on a section of skull approximately 80 mm by 50 mm in size, spanning the right frontal and temporal bones of a single specimen. Because of the strength of this mature skull, it remained stable even at the maximum pinning load of 100 lbs (444 N). So, for the fracture experiments presented here, the skin was removed and the skull was thinned over a ~10 mm diameter region using a surgical drill. The clamp was affixed with the single pin bearing in the center of this weakened region. This artificial thinning of the skull was intended to simulate conditions of reduced skull strength due to developmental immaturity, congenital thinness, degenerative disease, or prior trauma, but with the caveat that natural triphasic layering of the bone structure is clearly disrupted by this preparation in favor of bone primarily composed of the inner cortical layer. Nonetheless, the thinned skull is still comprised of bone that can be used to test the basic hypothesis of whether its time-dependent failure generates AE in such a manner that conforms to the critical state model (Eq. 1).

With the AE monitoring system active, the clamping of the skull began by placing the clamp after which the screw drive was tightened manually until the built-in indicator showed a load of 20 pounds. If the frequency and energy of AE events decreased or ceased, it was interpreted as evidence that the skull was stable under the current load. The load was then increased in 10 lb (222 N) increments by turning the screw drive.

Events generated by fracture were distinguished from those arising from clamp operation or other experimental noise using a filter based on waveform metrics. The details of this filter are described in Bunger et al. (in review). Here we will evaluate the events that are determined by the filter to be consistent with fracture, focusing on the period in the leadup to failure to determine of the acceleration of the events is consistent with Eq. (1) and if the time to failure is able to be predicted.

Here we consider one of the cases reported by Bunger et al. (2024a, 2024b, In review) along with a second case bearing similar characteristics. These are selected because they both exhibit failure that occurred in a time-dependent manner, that is, with a delay between the increase of the pinning load and the final moment of failure. In both cases, the failure was audible and accompanied by a clear and sudden displacement of the pin. Additionally, in both cases, the leadup period is observed to have an increase of AE event rate and amplitude, as shown in Figure 5. These were interpreted as foreshocks indicating impending failure by Bunger et al. (2024a, 2024b, in review). Here we analyze the AE energy data for these experiments to ascertain: 1) If there is a systematic cascade of these events that can be used to predict impending failure and estimate the time to failure, and 2) If other events that did not generate failure are indicative of stability. Both points are important, because a system designed to predict if a skull is progressing to failure under pinning loads should not only be effective at alerting to said failures, but avoiding false positives that would cause spurious disruption to a procedure.

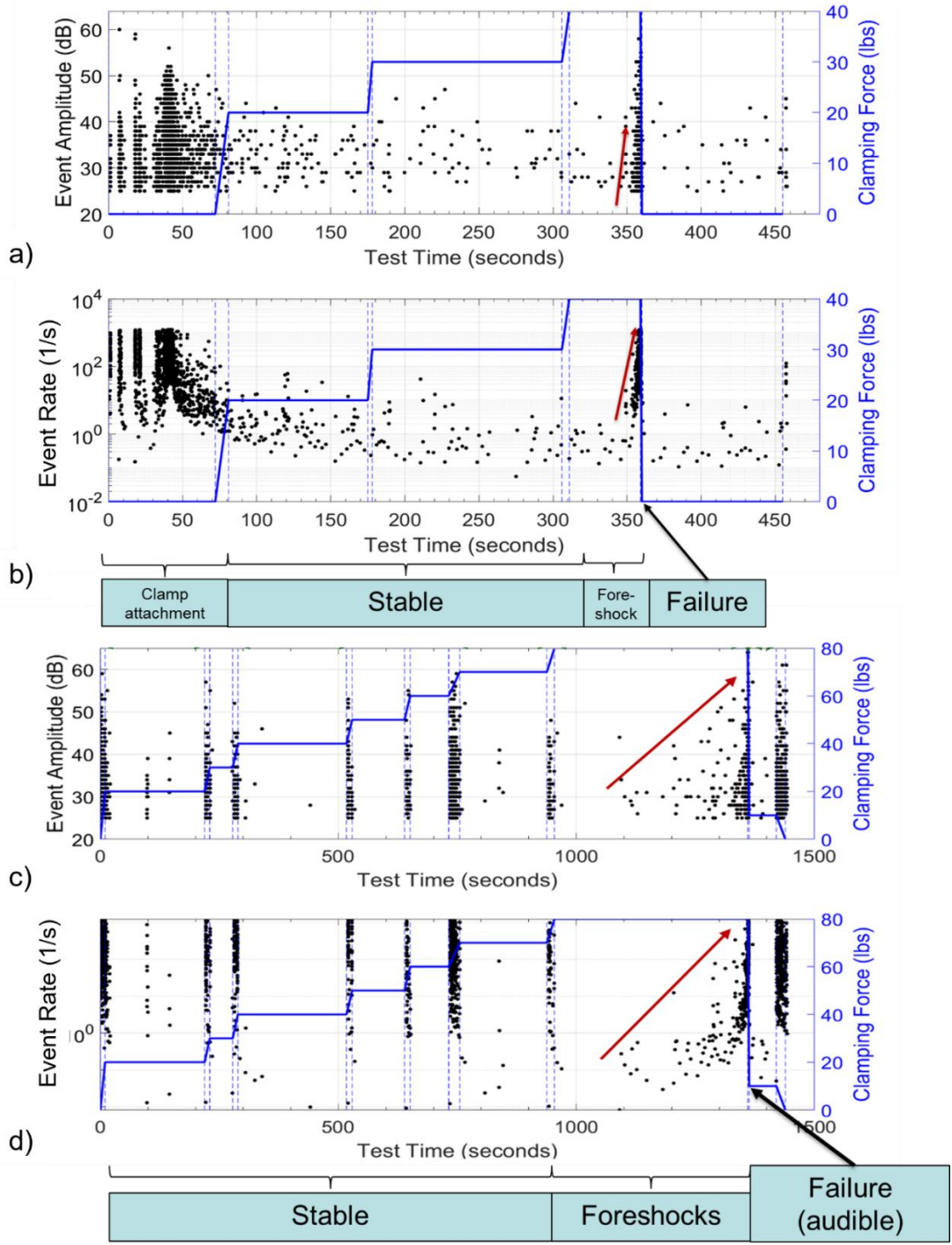

Figure 5: AE event rate and amplitude for Location 8 (a and b) and Location 3 (c and d), where data from Location 3 are redrawn based on material presented in Bunger et al. (2024a, 2024b, In Review). Annotations are added to indicate periods of time for behavior was apparently stable and then for which a foreshock sequence appeared to establish, with red arrows to highlight increasing event amplitudes and rates in leadup to failure.

## Analysis Methods

The measurable quantity $\Omega$ is taken as the cumulative AE energy. Other metrics were tried as a part of this study, including cumulative number of AE events and cumulative AE counts, but the results that gave the clearest conformance of Eq. (1) and were the most amenable to accurate prediction of failure time came from using the AE energy. Here the AE energy is computed by an order of magnitude estimate obtained by multiplying the waveform amplitude (in millivolts) by the waveform duration (in microseconds), as illustrated in Figure 6a. This cumulative AE energy was summed, for each event $i$, over all events $j \leq i$.

The resulting cumulative AE energy is therefore represented on an irregular time grid because each data point is at a time that an event occurred. The result is then normalized by the median value of the event energy. This is done so the energy rate will have the simple units of (1/s) and will represent the time elapsed to generate a release of energy commensurate with a median-sized event. Note one could also choose the mean, but because the event amplitudes are power law distributed (as shown by Bunger et al. 2024a, 2024b, In Review), the median is more appropriate as an indicator of the energy content of a "typical" event.

Having constructed a normalized cumulative AE energy function $\Omega$, the event rate $\dot{\Omega}$ is obtained using finite difference method with down sampling $f_d$, that is

$$i \leq f_d : \quad \dot{\Omega}(t_i) = \frac{\Omega(t_{i+f_d}) - \Omega(t_i)}{t_{i+f_d} - t_i}$$

$$f_d < i < N - f_d : \quad \dot{\Omega}(t_i) = \frac{\Omega(t_{i+f_d}) - \Omega(t_{i-f_d})}{t_{i+f_d} - t_{i-f_d}} \quad (3)$$

$$i \geq N - f_d : \quad \dot{\Omega}(t_i) = \frac{\Omega(t_i) - \Omega(t_{i-f_d})}{t_i - t_{i-f_d}}$$

Here *N* is the total number of events in the data set and the down sampling value $f_d$ is used to avoid high frequency oscillations in the derivative. The acceleration $\ddot{\Omega}$ is then found by the same finite difference scheme applied to the rate. For the granite beams, $f_d = 500$ except for the 35 second case, where $f_d = 50$ was used in light of the smaller number of recorded events compared to the other 5 cases. For the skull fracture cases, fewer events in total were detected and so $f_d = 20$ was used for these calculations. Note that many values were examined and it was found that these balanced resolving inflections with eliminating high frequency noise, and the ultimate result was not sensitive to the detail (i.e.

$f_d = 200$ and $f_d = 500$ for 5 granite beam tests gave ostensibly the same result, just with different apparent level of noise).

The analysis then entails both ascertaining onset of critical state by examining conformance of the data to Eq. (1) and using the behavior implied by Eq. (1) to predict the time of failure. Both of these follow Voight (1989). The beginning step is to divide both sides of Eq. (1) by the rate in order to obtain

$$\frac{\ddot{\Omega}}{\dot{\Omega}} = \eta \dot{\Omega}^{\alpha-1} \tag{4}$$

A linear and positive trend in a log-log crossplot of the left versus right hand side of Eq. (4) (as illustrated in Figure 6b) can be taken as indication of conformance to Eq (1) with $\alpha > 1$ and hence onset of criticality and the material entering, in the words of Voight (1989), the "terminal stages of failure".

To predict the timing of the failure, Eq (1) can be solved for the so-called "Inverse Rate", viz

$$\dot{\Omega}^{-1} = \kappa^{-1}(t_c - t)^{\xi}, \quad \kappa = (\xi/\eta)^{\xi}, \quad \xi = 1/(\alpha-1) \tag{5}$$

Approach of the inverse rate to zero as time approaches the critical time $t_c$ is indicative of the divergence of the rate to infinity and occurrence of catastrophic failure. Furthermore, for $\alpha \approx 2$ the inverse velocity decreases roughly linearly in approach to failure. The prediction of failure is therefore to select a window size, $c_{win}$, and generate a linear fit to the range $t_{i-c_{win}} \leq t \leq t_i$. This linear fit is projected to the x-axis (solving the resulting linear formula for x such that y=0) in order to determine the critical time $t_c$ (as illustrated in Figure 6c), which is the time at which the inverse rate becomes zero. For the granite beams the window size was taken as $c_{win} = 100$ except for the 35 second test, for which $c_{win} = 50$. For the skull fracture cases, $c_{win} = 5$. Once again, multiple values were examined and these were found to allow sufficient sensitivity to short term fluctuations in event rates without noise impeding the ability to interpret the results. Furthermore, the final results are relatively insensitive to the precise value chosen.

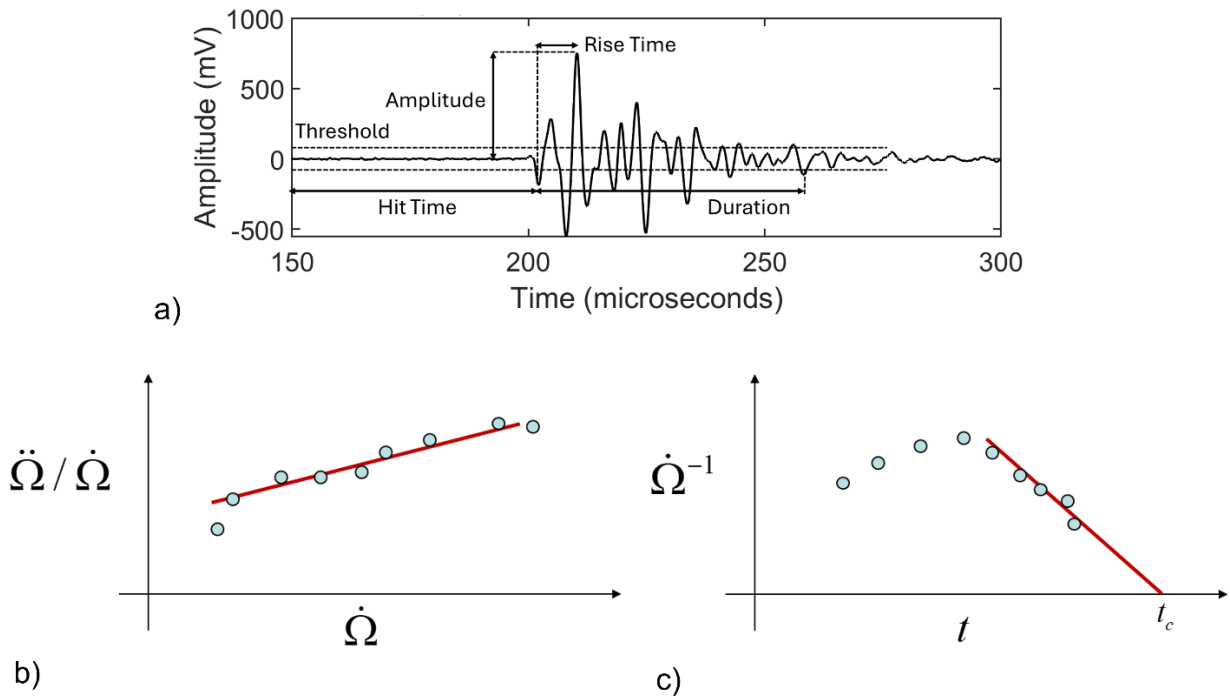

*Figure 6: a) Illustration of AE metrics, after MISTRAS manual [cite] but using an actual waveform generated by skull fracture, b) Illustration of log-log plot of the ratio of AE energy acceleration over rate versus AE energy rate, where a positive slope as sketched here would be taken to indicate terminal stages of failure, c) Inverse AE energy rate illustrating projection of quasi-linear cascade portion to the x-axis in order to predict the time of failure, $t_c$.*

## Granite Beam Results

AE energy rate and acceleration for all experiments using $f_d = 500$, the conformance to the critical state model is ascertained by plotting the ratio $\ddot{\Omega}/\dot{\Omega}$ versus $\dot{\Omega}$, as shown in Figure 7. The data from the beginning of the event rate acceleration until the time of failure (Phases III and IV from Figure 4) are coded as blue while other points are black. It is visually apparent that all six cases establish criticality during this period, which, again, comprises the last 30%-50% of their lifetimes. In some cases, the coefficient of determination, $R^2$, is very high (0.81-0.98, Figure 7d, e, f). In others it is not as large. However, even in two of these three cases where $R^2$ is smaller (Figure 7b, c), the linear positive trend in these plots is visually reasonable and $R^2$ is pulled lower due to intermittency that happened to occur in the AE events in these cases. That is to say, in these two cases the events were slightly more temporally clustered, leading to a behavior that is overall still consistent with critical state theory but with more points that lie off the trend. For Figure 7a, the test lasted only 35 seconds and produced thousands of events such that the event rate in the leadup to failure was so large that events were overlapping one another as they were being recorded, and hence two or more events were being counted as a single event. For this reason, the experiment portrayed in Figure 7a is most likely deviating from critical state theory due to difficulty detecting temporally overlapping events.

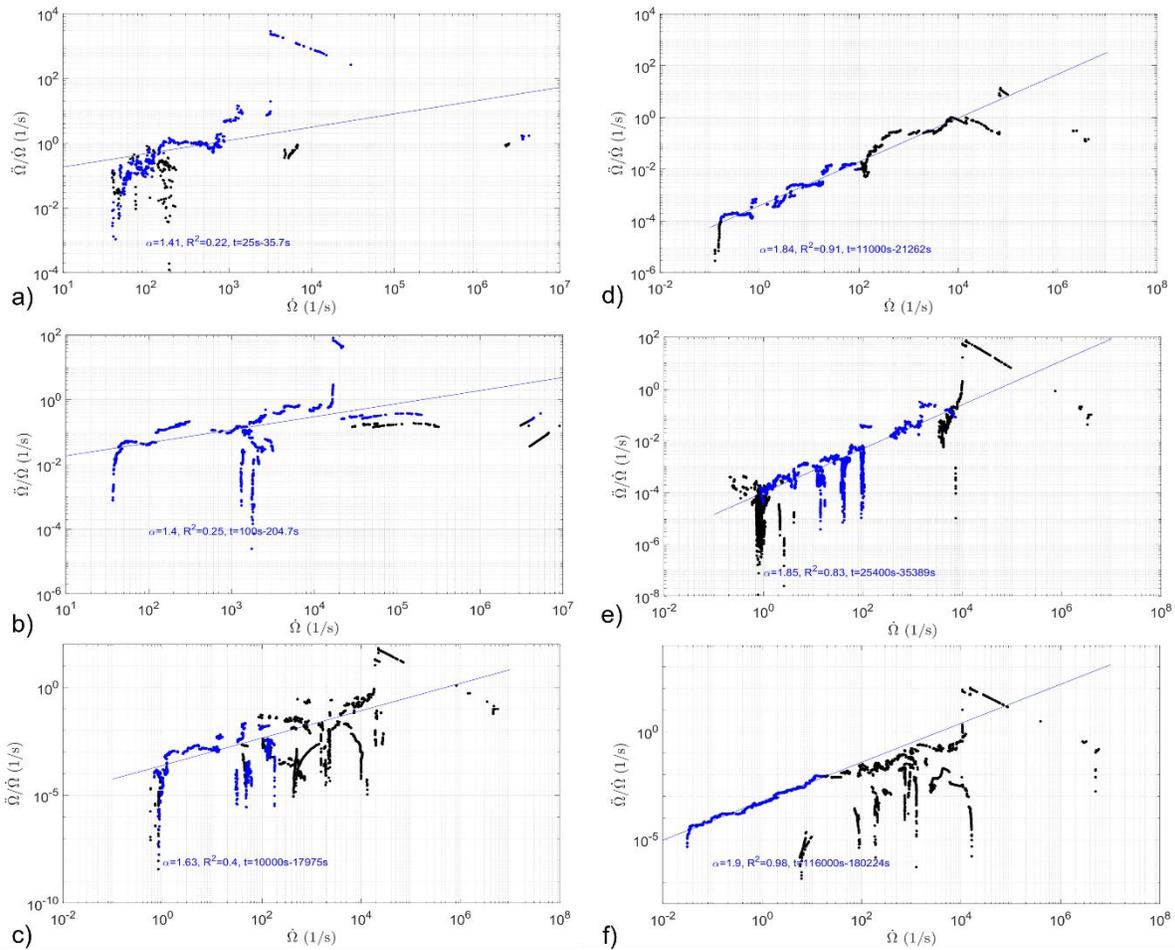

*Figure 7: Examination of criticality for 6 granite beam tests with approximate durations of: a) 0.5 min, b) 3.4 min, c) 5 hr, d) 6 hr, e) 10 hr, f) 50 hr. Here data points before the final cascade are shown as black which those during the final cascade are shown in blue.*

These results indicate that for all tests the criticality exponent is in the range $1.4 < \alpha < 1.9$. This confirms behavior consistent with progression to failure and indicates that the inverse velocity will be close to, but not precisely linear. We see that indeed this near linearity of the inverse velocity holds for each of the six experiments, as shown in Figure 8. In three of the cases (Figure 8a, e, f) a single slope prevails through the entire leadup to failure, while in the others there are changes of slope during the final stages of the specimens' lifetimes.

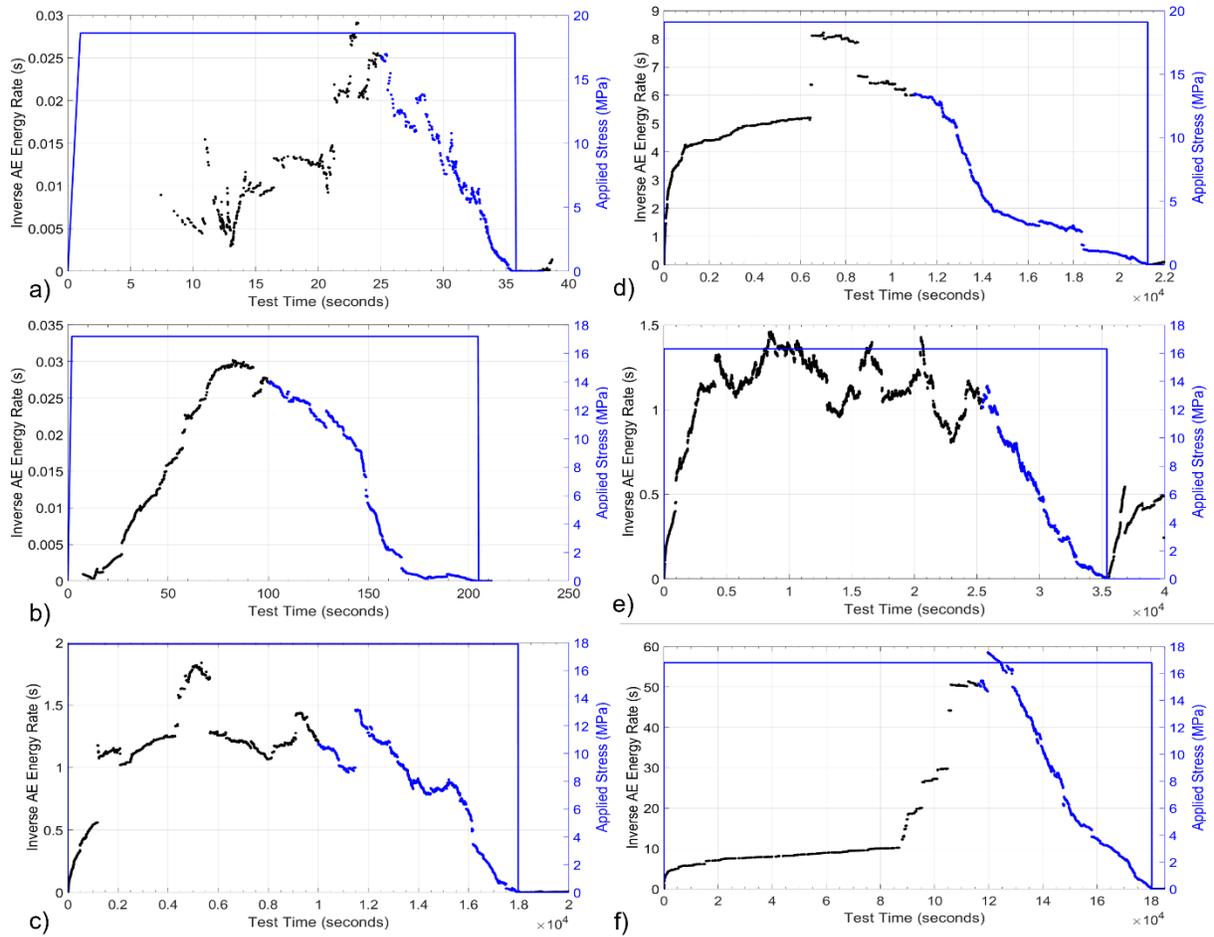

*Figure 8: Inverse AE energy rate for 6 granite beam tests with approximate durations of: a) 0.5 min, b) 3.4 min, c) 5 hr, d) 6 hr, e) 10 hr, f) 50 hr. Here data points before the final cascade are shown as black which those during the final cascade are shown in blue.*

The time to failure is then predicted for each case by generating linear fits to the inverse velocity for a moving window with a size of $c_{win}$ and projecting these to the x-axis to find a predicted failure time. For each data point, its current time is subtracted from this predicted time of failure to obtain estimates of the time to failure. In postprocessing, when the actual failure time is known, the actual time to failure can be plotted for comparison with the prediction. These comparisons are shown in Figure 9. In all cases, even those with changes in slope, the time to failure is effectively estimated during the final 30% of the specimen's lifetime. What this shows is that the slopes are of a similar order, and therefore even with the change, the order of magnitude of the estimated time to failure (which is what is seen in a plot with logarithmic scale on the y-axis, as in Figure 9) is unchanged. This is a rather striking result. For one illustrative example, the case that lasted over two days, the inverse of the AE energy rate, had it been carried out in real time, would have indicated progression to failure starting one day before failure actually occurred, and would have given an prediction that was accurate within one hour of the actual failure time for the final ~12 hours of the specimens lifetime.

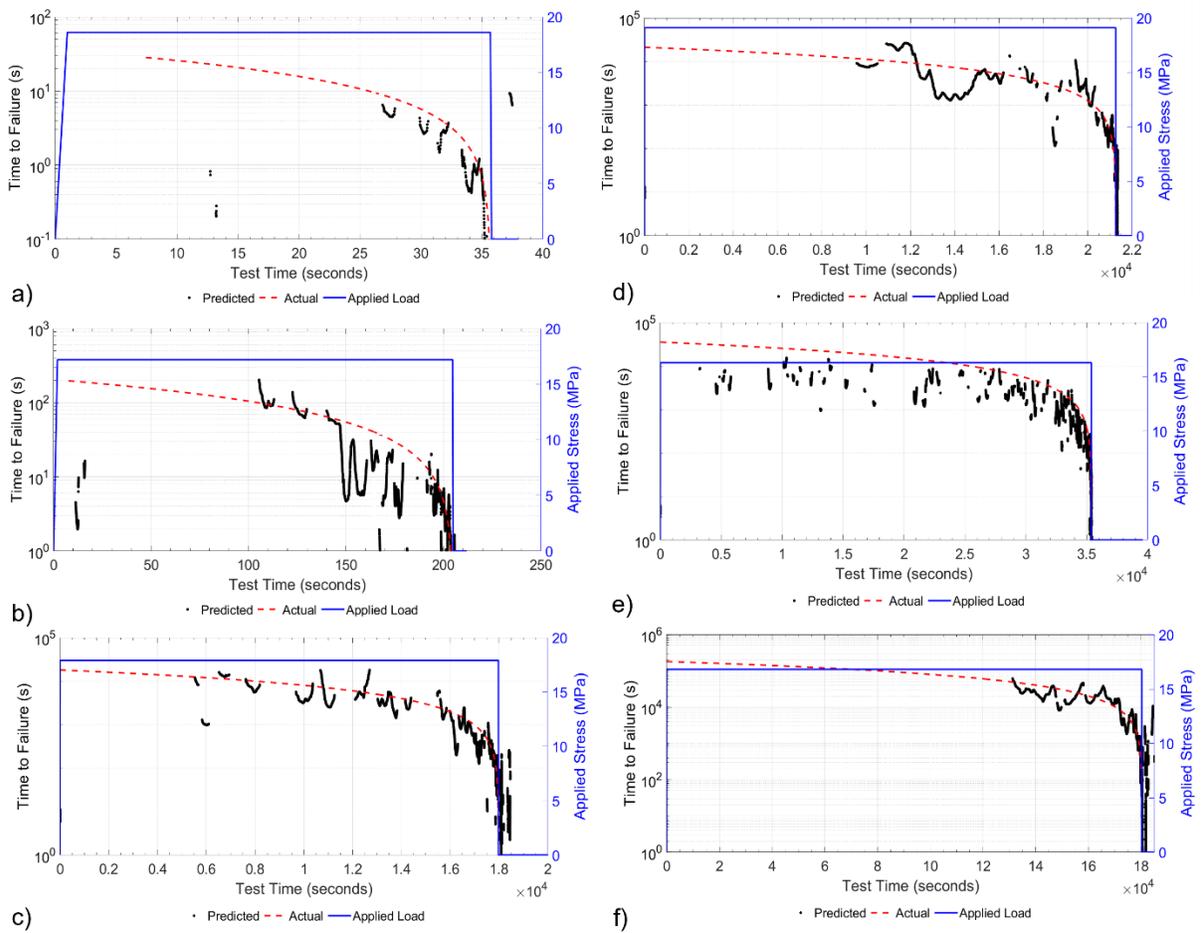

*Figure 9: Predicted versus actual time to failure for 6 granite beam tests with approximate durations of: a) 0.5 min, b) 3.4 min, c) 5 hr, d) 6 hr, e) 10 hr, f) 50 hr.*

Another striking result is that small bursts of AE energy that occur before the final progression to failure also have a slope that gives an estimate of the order of magnitude of the time to failure. Figure 10 gives the two cases where these initial cascades were most prevalent and shows they, too, provide reasonable order of magnitude estimates of the time to failure in spite of the fact that they are a part of a burst that apparently does not fully coalesce to failure.

Taken together, these results clearly indicate conformance of the granite beams to the rate-dependent failure model of Eq. (1) in the final (roughly) half of their lifetimes, with some long-range forecasting of time to failure available in some cases and accurate forecasting in all cased for the final 30% of the specimen's lifetime.

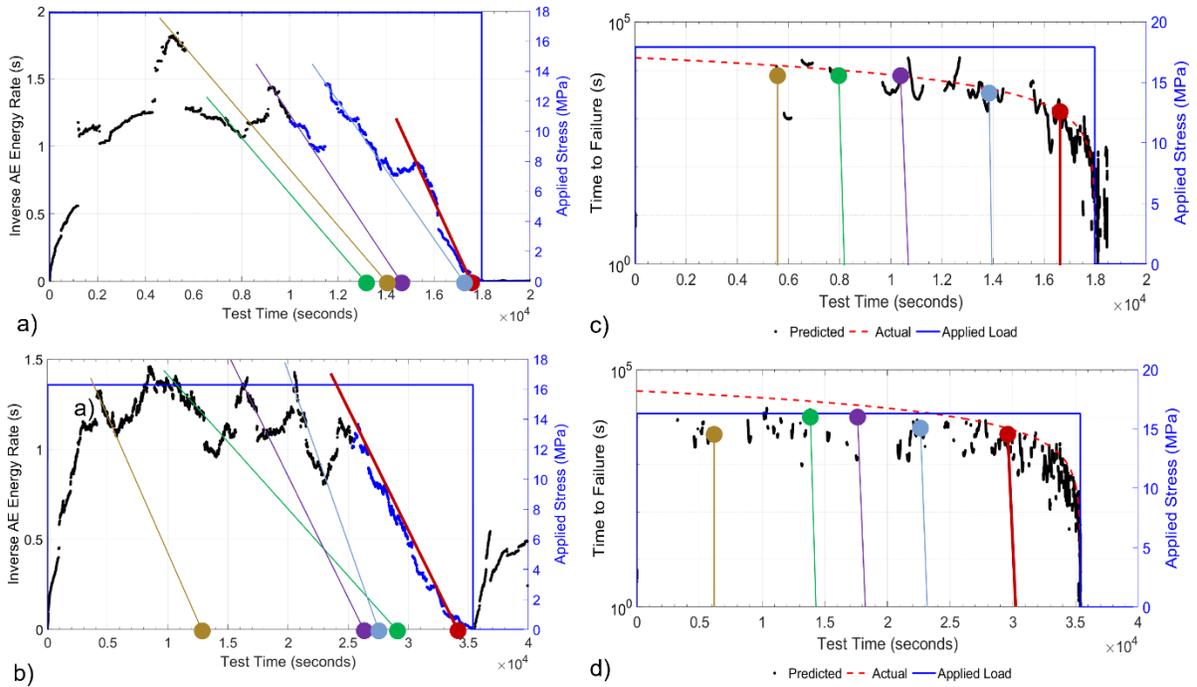

*Figure 10: Two cases illustrating early, non-terminal cascades of events that do not immediately progress to failure but that are still able to provide long-range order of magnitude estimates of time to failure even before the final stages of criticality establish. Here data points before the final cascade are shown as black which those during the final cascade are shown in blue. Subfigures a) and b) illustrate projections of these mini-cascades to the predicted failure time for the 5 hr and 10 hr cases, respectively. Subfigures c) and d) show the corresponding predicted time to failure at the moment of each mini-cascade.*

## Skull Fracture Results

The test at Location 8 generated AE after each of three increases in the pinning load to 20 lb (89 N), 30 lb (133 N), and 40 lb (178 N). A bit less than one minute after the increase to 40 lb (178 N) the skull failed and the pin was visibly seen to move rapidly inward when failure occurred. One can examine the acceleration to rate ratio, Figure 11a. The criticality in the leadup to failure is clearly indicated by the acceleration to rate ratio during the ~40 seconds before failure. The exponent implied is $\alpha = 1.89$, which is at the upper end of the range that was obtained for the granite beams and approaching the value of 2 that would imply perfectly linear decline of the AE energy rate during the terminal stages of failure.

Turning the attention to the inverse rate (Figure 12a), we see it progressing approximately linearly in the period after each load increase. However, it is increasing after the load changes to 20 lb (89 N) and 30 lb (133 N), which is indicative of $\alpha < 1$ and hence stability. After the load is increased to 40 lb (178 N), the behavior of the inverse rate changes immediately to linear decline, indicative of the terminal stages before material failure. It progresses with three periods that have differing slopes, but all of which indicate the imminence of failure.

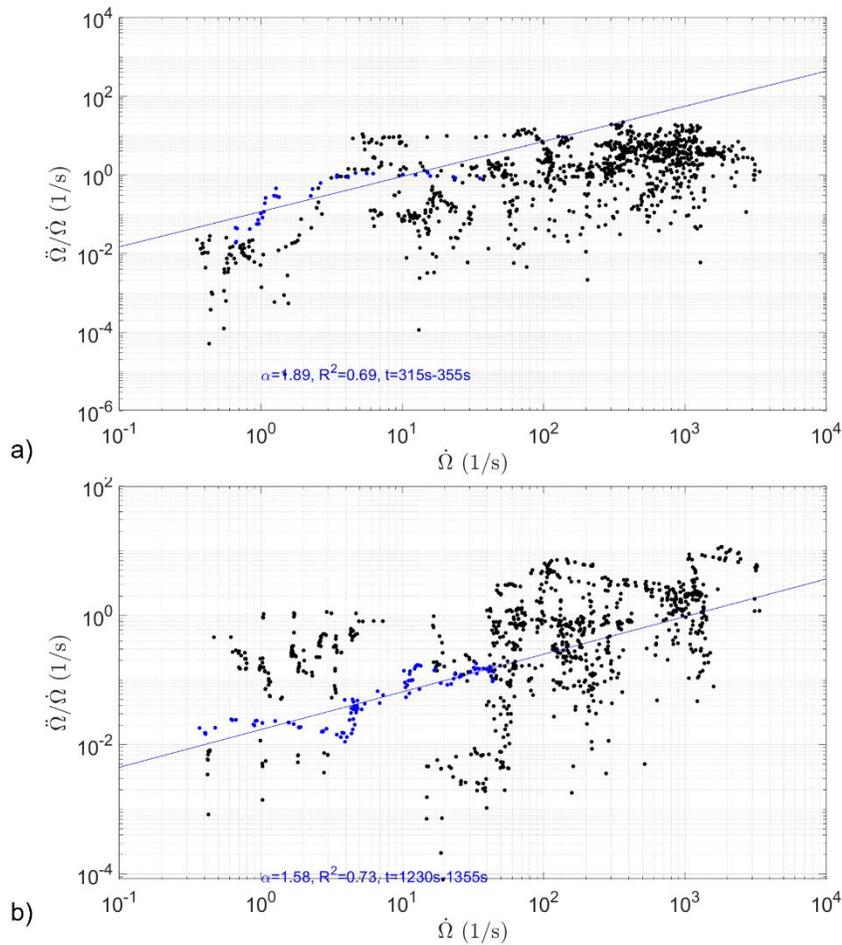

*Figure 11: Examination of criticality according to Eq X for skull fracture at: a) Location 8 during the period of 315-355 seconds, where failure occurred between 355 and 360 seconds, and b) Location 3, between 1230-1355s, where failure occurred between 1355-1360 seconds. The blue dots indicate events in the leadup to failure.*

Prediction of the time to failure is shown in Figure 12b to be reasonably accurate over the final 30 seconds. In practice, the establishment of the steady decline of the inverse rate would have caused an alarm within about 10-20 seconds after the load was increased to 40 lb (178 N) and therefore would have provided a surgical team with 30-40 seconds warning. This is typically sufficient time for the surgeon to receive an auditory warning, support the head, and turn the drive screw on the clamp to loosen it sufficiently to avoid fracture.

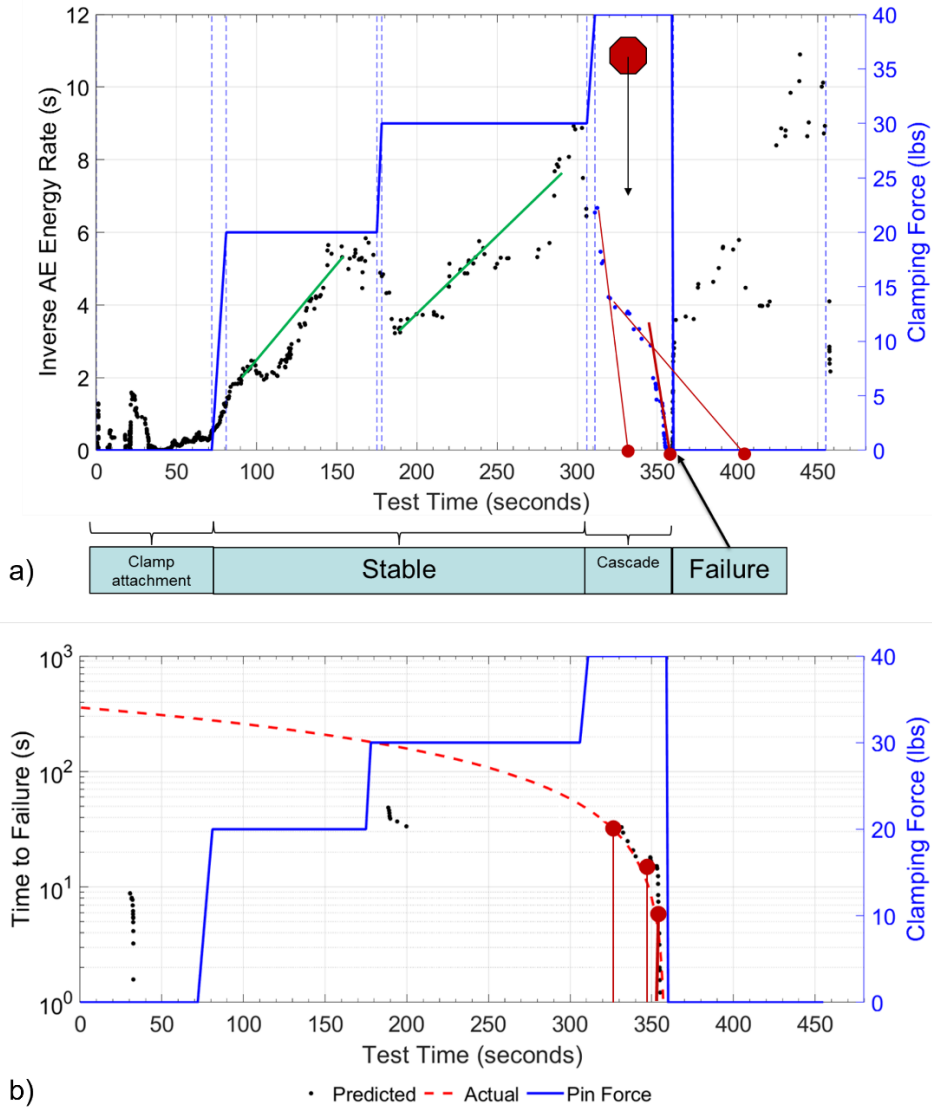

*Figure 12: a) Inverse AE Energy rate for skull fracture at Location 8. Stability is indicated by the increase of this quantity during the periods following increase of the pinning load to 20 lbs (89 N) and 30 lbs (133 N), where green lines with positive slope are included for illustrative purposes. After load is increased to 40 lbs (178 N) the AE Energy rate decreases indicating terminal stages of failure. The red octagon indicates the approximate time this cascade would have been sufficiently established to raise a warning to a surgical team. The three linear extrapolations show specific predictions of failure time that are in turn indicated in subfigure b. b) Predicted time to failure with actual time to failure for skull fracture Example 1. Three illustrative predictions are shown in correspondence with subfigure a.*

Similar behavior is shown at Location 3. The acceleration to rate ratio (Figure 11b) is shown to attain a linear form over the final 85 seconds before failure. During this period, the critical exponent is found to be $\alpha = 1.58$, which is a bit smaller than for Location 3 and in the middle of the range that was obtained for the granite beams.

The corresponding evolution of the inverse AE energy rate is shown in Figure 13a. In this case very little AE is generated until the pinning load is increased to 80 lbs (355 N).

However, AE was substantial during the ~400 seconds between the increase to 80 lbs (355 N) and the audible failure of the skull. Interestingly, the behavior for the first ~200 seconds was characterized by an increase in the inverse AE energy rate, indicating a stable period of around 50% of the lifetime of the skull at that load level, similar to the rock beams. After this period, the cascade of the inverse AE energy rate to the final failure establishes. The slope undergoes several changes, leading to moderate accuracy of the time to failure (Figure 13b). Nonetheless, once the first cascade established, a warning would have been generated (as illustrated in Figure 13a) giving, in this case, around 70 seconds during which the surgical team could have taken action to avoid patient injury.

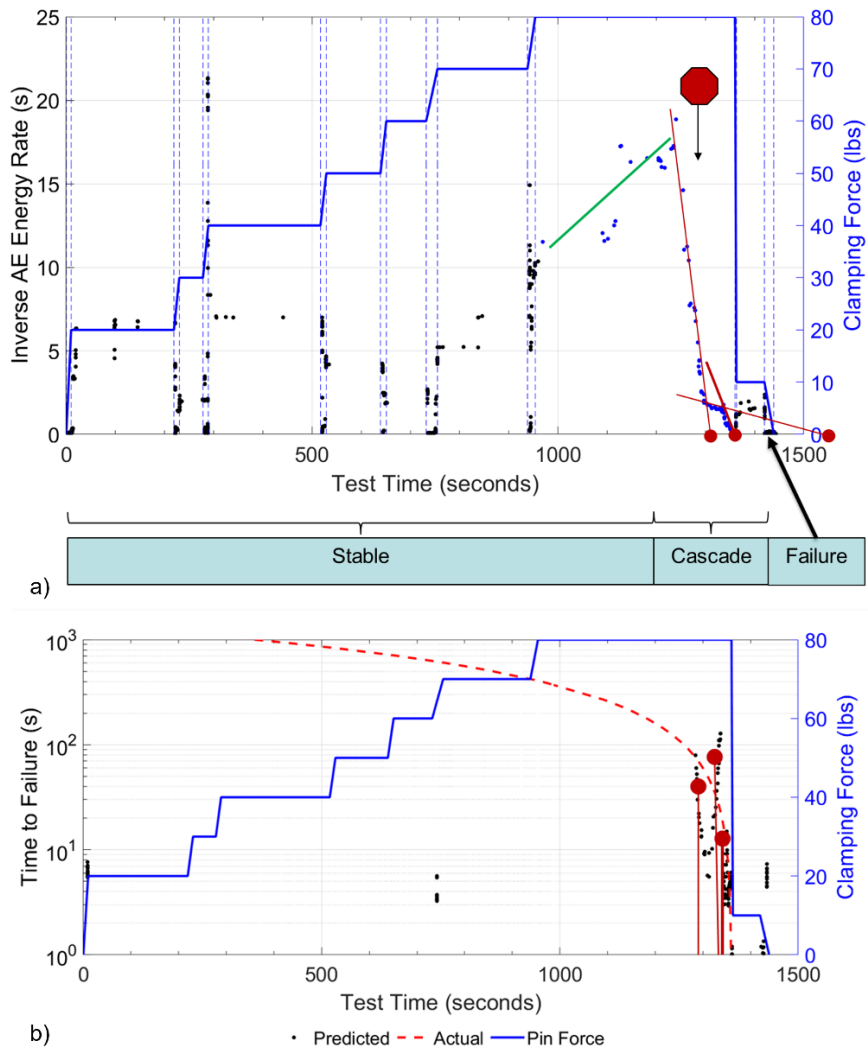

*Figure 13: a) Inverse AE Energy rate for skull fracture at Location 3. Little AE is generated until load is increased to 80 lbs (355 N). Stability is indicated initially after increase to 80 lbs (355 N) by the increase of inverse AE Energy rate (green line with positive slope is included for illustrative purposes). After initial stability the terminal stages of failure are indicated by decrease of inverse AE Energy rate. The red octagon indicates the approximate time this cascade would have been sufficiently established to raise a warning to a surgical team. The three linear extrapolations show specific predictions of failure time that are in turn indicated in subfigure b. b) Predicted time to failure with actual time to failure for skull fracture Example 1. Three illustrative predictions are shown in correspondence with subfigure a.*

## Discussion

The critical state model of Voight (1988, 1989), Eq. (1), provides a mathematical framework for predicting time to failure based on measurements. It is a manner of divergence of the measurable quantity to infinity that differs from other growth laws, notably the exponential growth model. In the past there have been implications that seismic and/or acoustic emission precursors to failure grow exponentially in the leadup to failure (e.g. Sammonds et al. 1992, Stanchits et al. 2006, and original model proposed in Winner et al. 2018). The analysis presented here shows that AE generated by both rock and bone in the leadup to failure is diverging as a power law, with a well-defined critical time for catastrophic failure. This contrasts with exponential growth, which approaches infinity as time goes to infinity and does not admit a cascade that can be used to predict time to failure.

There is another important contrast between the exponential and critical state models. The reason for ubiquity of exponential growth models is clear; for many processes, the rate of growth of a quantity is proportional to the quantity itself. The fundamental reason for the widespread applicability of the critical state model, on the other hand, is not as clear. Such understanding would be useful for interpretation of data, especially in a study like this one where there is apparent commonality in behavior between two materials that are ostensibly different in many ways.

One helpful starting point is perhaps the earliest and most well-known justification of Eq. (1) (Varnes 1989, Kilburn and Voight 1998, Main 1999, Main 2000), which shows that a subcritical crack growth model can lead to a divergence of the crack length in the form of Eq. (2). We will return to this specific example, but first, taking a more general approach, consider $Y$ to be a response (such as a displacement, crack length, strain, AE energy rate, etc.) to a driver $X$ (a force, stress, stress intensity factor, etc.). Then, let the rate of change of the response be a power law of the driver

$$\frac{dY}{dt} = A\left(\frac{X}{X_0}\right)^n \tag{6}$$

In turn, let the driver be a power law of the accumulated response

$$X = BY^m \tag{7}$$

This feedback loop can be shown to generate the critical state law (Eq. 1) by taking a derivative of both sides and simplifying to obtain

$$\frac{d^2Y}{dt^2} = \eta\left(\frac{dY}{dt}\right)^\alpha, \qquad \alpha = \frac{nm-1}{nm}+1, \qquad \eta = nm\left(\frac{X_o^n}{AB^n}\right)^{\frac{nm-1}{nm}-1} \qquad (8)$$

This analysis shows that the condition for instability, $\alpha > 1$, requires simply that the product of the exponents $nm > 1$. Furthermore, for $nm \gg 1$, $\alpha \to 2$. So, taking the classical approach that combines Charles' law (Charles 1958) for the velocity of a crack under subcritical conditions with an LEFM based coupling between the crack length (*L*) and the stress intensity factor (*K*), that is

$$\frac{dL}{dt} = A\left(\frac{K}{K_c}\right)^n, \quad K = B\sigma L^m \qquad (9)$$

Here $K_c$ is the fracture toughness, $\sigma$ is the applied stress, and for a straight crack subjected to remote tension, the exponent *m*=1/2. The coefficient *A* and exponent *n* are experimentally-determined, and for many materials *n* is in the range of 20-50 (Main 2000, Atkinson and Meredith 1987). It has been found for the Cold Spring Charcoal Granite used in the experiments of Winner et al. (2018) to be *n*~27 (Lu et al. 2020). Because of this large range of typical values of n, subcritical crack growth is expected to give $1.9 < \alpha < 2.0$. Additionally, for human cortical bone it has been found in the range 20-40 (Nalla et al. 2005), which would imply a similar range of expected $\alpha$.

Another illustrative example comes from a rate-dependent material model having the form (Borja and Kavazanjian 1985, Hickman and Gutierrez 2007, Perzyna 1966, Taylor 1948)

$$\frac{de}{dt} = A\left(\frac{\sigma}{\sigma_0}\right)^n \qquad (10)$$

Here *e* is the viscoplastic strain, $\sigma$ is the applied stress, $\sigma_o$ is a reference stress, and *A* and *n* are experimentally determined. Note that *A* represents the strain rate when $\sigma = \sigma_o$ and the exponent *n* varies widely. For rocks, Hashiba and Fukui (2015) summarize and extensive literature showing *n* ranging from 28-115, but typically in the range 30-70. For bone the values tend to be smaller. For example, Qiu et al. (2023 especially Figure 8 therein) find *n* in the range 6-76 for cortical bones, with the formalin fixed specimens occupying the lower end of the range relative to fresh and dehydrated specimens. For hydrated lamellar bone Peruzzi et al. (2021, see Section 4.2 therein) find *n* in the range 17-24. Finally, for fresh skull bone, Zhai et al. (2020, see Figure 9 therein) show strain rate dependence that can be readily fit with Eq. (10) to estimate the range for *n* as 5-10. In any case, a critical feedback loop can form if the stress is coupled to the accumulated strain, which is in general dependent upon loading configuration, as

$$\sigma = Be^m \qquad (11)$$

such that the exponent $m > 1/n$.

For rock, bone, and other quasi-brittle materials, an important commonality is failure occurring due to microcrack coalesce. Indeed, the methods for predicting catastrophic failure from AE energy rates presented in this paper rely on this mechanism. A number of microcrack coalescence models exist in the literature (e.g. Huq et al. 2019), but the direct connection to a rate-dependent, critical state failure model is yet to be clearly established and is recommended as future work.

In summary, the broad applicability of the critical state law (Eq. 1) to material failure (and beyond) is because of the widespread occurrence of materials (and systems in general) for which a response rate is a power law of a driver and the driver is, in turn, a power law of the accumulated response. Additionally, the tendency of the exponent $\alpha \to 2$ arises because of the manner in which the exponents of the two coupled laws, *n* and *m*, combine when the laws are coupled together. As a result, the cascade to failure of the inverse response rate, which is used to predict time to failure, is nearly linear for many materials and systems.

As a final practical point, let us consider the procedural implications of a warning system of skull failure that gives 30-60 seconds for the surgical team to take action to avoid patient injury. A viable operative setup would include the feedback from the AE detector being displayed to an ancillary staff member stationed in the room already tasked with similar duties such as the neurophysiologist or anesthesiologist. Alternatively, the system could be unmanned and emit a specific auditory tone when a failure cascade is detected. In either case, the surgeon would be alerted of an imminent fracture. A basic rapid response maneuver would involve reaching through or under the drapes to support the head while simultaneously loosening the head clamp manually until the cascade is aborted. At this point. The drapes could be removed and the patient taken out of pins and repined or, if deemed safe, left in pins at the lower pin force that resulted in the termination of the cascade. Loosening the pins takes approximately 1-2 seconds per 20 lbs of force. Hence, a 30 second warning window gives the surgeon 28 seconds to understand the warning indication and to then secure the head. As the surgeon is typically physically positioned at the patient's head, a total time to execute a response would be unlikely to exceed 20 seconds, which provides an ample margin for unanticipated factors.

# Conclusions

The evolution of the AE energy rate is explored for both rock and bone with failure progressing under constant loading conditions. Time dependent failure of rock is examined through analysis of data obtained for granite beams subjected fixed loading under three point bending and monitoring the time-dependent failure. Six experiments under differing load levels resulted in failure times ranging from 35 seconds to two days. It was previously observed that the AE event rate declines for the first roughly 40% of the lifetime of the specimen. It then slowly accelerates in the lead up to a final burst of events before failure. Here we show the slow and rapid acceleration stages are consistent with a rate dependent material failure law applied to the AE energy. One consequence is a quasi-linear decrease of the inverse AE energy rate that lends a straightforward prediction of time to failure. This prediction is typically accurate over the final third of a specimen's lifetime. Additionally, earlier cascades of AE energy that are apparently arrested also provide long range, order of magnitude forecasts of time to failure.

From the commonality of bone and rock as quasi-brittle materials that fail by microcrack coalescence, we then hypothesize that the rate dependent material failure law will also govern time dependent skull fracture. This is tested experimentally with human cadaver skulls that are loaded under pinning loads similar to those used for immobilization of the head for neurosurgical procedures. AE energy rates are shown to be consistent with the rate dependent material failure law including a quasi-linear cascade of the inverse AE energy rate in the lead up to failure. This cascade provides a distinctive warning of impending failure, enabling warning at 30-70 seconds before failure. This is a sufficient timeframe for a surgeon to pause the procedure, support the head, and loosen the clamp in order to halt the failure progression and avoid patient injury.

Finally, we return to the origin of catastrophic failure of rate-dependent materials. The failure cascade is shown to be a result of an underlying material behavior wherein the rate of a response is a power law of a driving force and the driving force is, in turn, a power law of the accumulated response. Considering rate dependent failure in this way explains its widespread applicability and its tendence for the exponent of the failure law to tend to 2 and hence the inverse rate of the response of the system to go to zero with a linear trend. This illuminates the basic ingredients of models developed both in the past and future that should be present if the model is to be capable of capturing the catastrophic cascade to failure and the predictability of the time to failure during terminal stages of the failure progression.

## Competing Interests

APB and MM are coinventors on Patent Cooperation Treaty (PCT) filed as:  Bunger AP, Hager CC, Hartz OG, Harbert W, McDowell MM. Determining a State Arising During a Medical Procedure Involving Bone Via Monitoring Acoustic Emissions from the Bone. Provisional

Patent Application No. 63/662,125 filed on 1 October 2024, Now PCT Application PCT/US2025/034360.

# Acknowledgement

Partial support for this work was provided via the R.K. Mellon Faculty Fellowship in Energy (AB). Additional support was provided by the University of Pittsburgh, in particular via Research Development Funds (AB). Charles Hager provided technical support for developing and deploying the instrumented pin and AE detection system and Kyle Affolter provided technical support for preparation of the cadaver sample. Early discussions of geophysical implications for detecting bone fracture with Professor William Harbert are also acknowledged.

# Data Availability

Data for Skull Location 3 and supplementary information about data analysis can be obtained from http://d-scholarship.pitt.edu/id/eprint/46998 . Data for Skull Location 8 and the Rock Beams can be obtained from http://d-scholarship.pitt.edu/id/eprint/48634.